\documentstyle[preprint,version2,aps]{revtex}

\begin{document}


\newpage
\begin{title}
An Interactive NeXTSTEP Interface to a Fortran Code\\
for Solving Coupled Differential Equations
\end{title}

\vskip 0.1in

\author{Richard R. Silbar}
\begin{instit}
Theoretical Division, Los Alamos National Laboratory\\
University of California, Los Alamos, New Mexico 87545
\end{instit}
\receipt{August XX, 1992}

\vskip 0.2 in
\begin{abstract}

This paper describes a user-friendly frontend to a Fortran program that
integrates coupled nonlinear ordinary differential equations.  The user
interface is built using the NeXTSTEP Interface Builder, together with
a public-domain graphical palette for displaying intermediate and final
results.

The main obstacle was implementing communication between the Objective-C
environment governing the NeXTSTEP interface and the Fortran code.

This was overcome by breaking up
the Fortran into separate subroutines (which compile as C function calls),
corresponding to the various control buttons in the interface.

In running the code for a given set of equation parameters the user sees

a plot of the solutions at each stage of the iterative process.
In the case of a successful sequence of iterations, the initially

discontinuous curves smooth out as the scale parameters of the

solutions are adjusted to achieve a solution to the nonlinear equations.
If the iterative process goes astray, as it often does for a poor choice

of starting scale parameters, the user has the opportunity to stop and

start over with a better choice, guided by the result of the previous run.
The ease of interaction with the equations also allows the user to develop
an intuition regarding their solutions and to explore the parameter space

for the equations much more quickly.

\end{abstract}

\newpage


\section{Introduction and Background}

In the course of my research as a nuclear physicist I recently had occasion to
write a Fortran code for solving coupled, nonlinear ordinary
differential equations (ODEs).
The code, which originally ran on a VAX, used appropriate Numerical
Recipes\cite{NumRep} to carry out a Runge-Kutta integration of the
equations.
This paper describes how I came to develop a user-friendly graphical

interface to this code.

The interface makes it much easier to use the code for finding the
solutions to the nonlinear ODEs.

In particular, the original Fortran code solves second-order

coupled ODEs for two functions, $F(r)$ and $G(r)$, by matching at some
intermediate distance.
The boundary conditions at the origin are known and the equations themselves
determine indicial behaviors near $r=0$.
There are also desired asymptotic behaviors at large distances

(exponential falloffs).
However, one doesn't initially  know the sizes of derivatives at the origin

or the asymptotic normalizations, since these are determined by the

nonlinearities of the equations.
So, what the user of the code does is to choose these four scale parameters
arbitrarily for a first pass through the equations.
These parameters are then subsequently refined by
successive iterations until the equations are well-satisfied.\cite{proc}

This Fortran program was invoked as a batch process from a

(text only) terminal window.
The user could semi-interactively input parameters, {i.e.}, the code would use
default values and, in the terminal window, query the user

as to whether he or she wanted to make changes.\cite{inpfile}
This mode of operating was tolerable when running the code on a VAX mainframe,
 but the problem with this procedure is that non-linear equations are tricky.
Much of the time a choice of starting scale parameters ends up making
a disastrous step, which wastes the whole batch run.

Because of a change of station, I soon had to transfer my Fortran
computing from the VAX to a NeXT workstation.
Consideration of cost led me to use the public domain ``compiler" f2c, which
translates Fortran code into C code, and then compiles that with the
Gnu cc compiler that comes bundled with the NeXT operating system.\cite{f2c}
This, as will be seen, turned out to be a serendipitous, as well as economic,

choice.

After the move to the NeXT, however, some things about running the program

as a batch job became annoying.
For example, performing another calculation meant starting a new job
and re-entering all the input parameters.
Another annoyance was the graphing of the resulting solutions.
On the VAX, plots of solutions could be made automatically by having the
Fortran code write out graphics files to the terminal running in a
Tektronix 4014 emulator mode.
On the NeXT, however, the plotting had to be done by post-processing an output
file after each run.
I began looking for a better way to work.

Perhaps more of a reason to make a change came from the above-mentioned
difficulties of finding solutions to the nonlinear equations themselves.
The code often cannot find reasonable
Newton-Raphson corrections to the scale parameters
if the initial values are poorly chosen

(i.e., too far from the solution values).
If one could immediately see graphs of the results of the first pass,
that would help in making a good initial choice.
(If it's very bad, one would not proceed with that run and try another choice.)
Likewise, seeing graphs of intermediate pass results would allow the user
to opt out of a calculation that starts to ``blow up''.

Thus, it soon became obvious that I should try putting a interactive
interface on the Fortran ODE solver program.
However, the NeXTSTEP operating system and its graphical user interface (GUI)
are based on Objective-C, an object-oriented extension of the C programming
language.
So the basic question now became how to get Objective-C to talk to

Fortran (and vice versa)?

\bigskip
\section{To Prove the Concept}

To answer that question, consider a simple ``Hello, Fortran!'' application,
which has the window interface shown in Fig.\ 1 and which
does the following:
\begin{itemize}
\item{
The user inputs (clicks, then types) a number into a small text window, a
FormCell object labeled ``In'', in an on-screen window representing the
``Hello, Fortran!'' application.}
\item{
Clicking on a Run button in that window reads that number and
sends it to a Fortran program.}
\item{
The Fortran code multiplies it by 7 and displays the ``answer'' in another
FormCell object (labeled ``Out'') in the window.}
\end{itemize}
\medskip

To build such a program as a NeXTSTEP application one first builds the
window using the Interface Builder\cite{NeXTSTEP} (IB), a developer's toolkit
that comes with every NeXT workstation having the extended
distribution.
How the IB is used will only be described briefly in this paper;

the reader is referred to Ref.\ 3 for details.

After launching the Interface Builder, one uses the mouse to drag in and drop

various graphical objects from the IB palette in the application's window.
For the ``Hello, Fortran!'' application it is only necessary to add two Forms
(both one cell only) for input and output and a Run button.

Now one creates a custom class, call it {\tt RunFortObject}, by subclassing the
generic {\tt Object} class.
This object will contain all the specialized behavior we need to control the
graphical interface window and access the Fortran module.
At the time {\tt RunFortObject} is made, the IB also creates an {\it instance}
of it, {\tt Run\-Fort\-Object\-Instance}.
In this class one declares the existence of two outlets, {\tt input\-Form} and
{\tt output\-Form}, and one method, {\tt run\-Fort\-Method:sender}.
(A method is the object-oriented equivalent to a function call;
the argument ``sender'' will refer to the Run button object.)

The {\tt Run\-Fort\-Object\-Instance}'s outlets,
{\tt input\-Form} and {\tt output\-Form}, now need to be specified,
i.e., the {\tt Run\-Fort\-Object\-Instance} object must know where those

graphical objects are on the screen.
Likewise, the Run button must know what object to tell that
it has been clicked and what message should be sent there.
These connections are made by dragging a connecting line,
again using the mouse, from the graphical objects in the window
to their corresponding target objects.
In this way code in the {\tt RunFortObject} which refers to, say,
the {\tt output\-Form} will know where to send the Fortran answer for display.
Similarly, the Run button is connected to the
{\tt Run\-Fort\-Object\-Instance}
as target and will trigger the action coded in {\tt run\-Fort\-Method}.

Finally, by clicking on the Parse menu item, the Interface Builder creates a

class header file, {\tt run\-Fort\-Object.h}, which {\it defines} the

interface (see Fig.\ 2).
The IB also creates a shell of the Objective-C that needs to be

filled in for the class
implementation file, {\tt run\-Fort\-Object.m}.
Nothing further has to be done with the header file (in this case), but
the {\tt runFortObject.m} file needs fleshing out.
The following material was added, by hand, to {\tt runFortMethod}
(see Fig.\ 3):
\begin{itemize}
\item{
  First, read the {\tt inputForm}, write its value to a file, {\tt input.data},
  and close the file.}
\item{
  Call the Fortran code, which is actually the subroutine {\tt hellosub} in the
  {\tt hellosub.f} file (see Fig.\ 4).
  The compiler-translator, f2c, creates a C function, {\tt hellosub\_()},

  corresponding to this
  subroutine which can then be called by that name in the
  Objective-C code.}
\item{
  The Fortran subroutine opens and reads the input data from {\tt input.data}
  and massages it (multiplies it by 7 in the present case).
  It then writes its output to the file {\tt output.data},
  closes that file, and returns.}
\item{
  The Objective-C routine, {\tt runFortMethod}, continues by opening
  the {\tt output.data} file, reads it, and
  finally displays the number there in {\tt output\-Form},
  {i.e.}, in the main window.}
\end{itemize}

Three details in this code should be noted.
First, Fortran is case-insensitive, but Unix filenames do care about case.
Second, the codes in Figs.\ 2 and 3 include {\tt printf}'s and {\tt print}'s,
seen here as commented out, for debugging purposes.
If the application is launched from a terminal window (a shell), these debug
printouts would appear there.
The ability to print out intermediate results turned out to be very useful for
debugging.
Finally, it is necessary to include the {\tt \#import} statements for the
header files {\tt appkit/Form.h} and {\tt stdio.h}, so that the C compiler will
be able to find definitions and other necessary things.

The final step is to ``make'' the application, to test it, and to iterate as
necessary.
The Interface Builder generates a standard Unix Makefile, which is not to be

touched.
However, the code developer can customize that Makefile as needed by creating
{\tt make.preamble} and {\tt make.postamble} files, which are read and executed
by the standard Makefile.
In this particular application one needs to declare {\tt OTHER\_OFILES} to
force recompilation of {\tt hellosub.f}, if it is changed (Fig.\ 5).
Also, to be able to load the Fortran I/O routines, etc., one must declare
the f2c library.
(Alternatively, one could declare the library by adding it to the ``other

libs''category in the Project Inspector window of the IB.)

Such a NeXTSTEP application does work, and, as a learning tool,
it provided the clues for how to front\-end a more ambitious Fortran code,
such as the ODE-solver.  Details of how this was done are discussed

in the next section.

\bigskip
\section{The frontend for the coupled ODE solver code}

The {\tt rhoSky} application solves coupled nonlinear ODEs for
two functions, $F(r)$ and $G(r)$.
These functions involve four as-yet undetermined scale parameters,

$A$, $B$, $C$, and $D$, which will be fixed by the iterative process
we employ to solve the nonlinear equations.
Details regarding these equations and some of the physics behind them
are given in an Appendix.

The Main Window for the {\tt rhoSky} application is assembled using the
Interface Builder as in the ``Hello, Fortran!" example discussed above, but it
involves more objects (Fig.\ 6):
\begin{itemize}
\item{
  An input Form object for various input parameters:

  coupling constants of the Lagrangian;
  particle masses;

  the matching radius $r_F \equiv {\tt xf}$;

  the starting point in the asymptotic region,

  $r_2 \equiv {\tt x2}$, for the backward integration;
  the Runge-Kutta precision parameter, $\epsilon$;
  and the boundary value of $F(0)$ (which is usually taken to be $\pi$,
  corresponding to baryon number $B=1$).}
\item{
  An input Form for the starting values of the four scale parameters ($A$, $B$,
  $C$, $D$) which are adjusted to make $F(r)$, $G(r)$ continuous and smooth at
  $r_F$.}
\item{
  A Form for displaying the present values of the adjusted scale parameters,
  i.e., what was used in the present iteration that is displayed in the
  Plot Window.}
\item{
  A Form for displaying the discontinuities of $F(r)$ and $G(r)$ and their
  first derivatives, $F(1) \ldots F(4)$, at the match point and the
  calculated corrections, $\Delta A \ldots \Delta D$, to the scale

  parameters based on these discontinuities.
  (A warning: these discontinuities $F(i)$
  are not to be confused with the solution $F(r)$.)}
\item{
  An output Form for displaying post-processed computations, such as the
  Skyrmion mass, {\tt M\_sky}, that depend on the solutions $F(r)$ and $G(r)$.}
\item{
  Default values in the Form Cells for input and scale parameters;
  these can be changed
  by the user by clicking on the desired cells and editing them,
  before clicking on Run.}
\item{
Various control buttons:\\
\indent
  {\bf Run}---does a Clear and performs first the pass (i.e., a first

  iteration) of a new calculation, with input and scale parameters as

  displayed.\\
\indent
  {\bf Continue}---goes on to next iteration, if desirable.\\
\indent
  {\bf Finish Up}---writes out the final solution to file {\tt output.data}
  and performs and displays post-processing computations.\\
\indent
  {\bf Clear}---clears the Plot Window, discussed below, and the output
  form cells.
}
\end{itemize}
\noindent
The evolution of the Main Window as one goes through the steps to find a
solution is shown in Fig.\ 5.

In addition, the application also has a Plot Window for displaying the

present (or final) calculated $F(r)$ and $G(r)$ (Fig.\  \ref{plotwin}).
The code for plotting in this window is a custom-built object,
{\tt nxyPalette},\cite{nxyPal}, available in the public domain.
{\tt nxyPalette} in turn is based upon a NeXTSTEP plotting application,

{\tt nxyPlot}, also in the public domain.\cite{nxyPlot}
The expanded {\tt RunFortObject.m} file, discussed below, includes three
methods needed for plotting (besides those methods already in the {\tt nxyView}
object provided by {\tt nxyPalette}).
The graph in the Plot Window is updated after every pass (Run, Continue, or
Finish Up).
As shown in Fig. \ref{plotwin}, the user sees how the curves,

which are initially discontinuous, smooth out as a solution is achieved.

The {\tt RunFortObject.m} file is now more complex and
contains eight methods.
Four of these are connected to buttons:
\begin{itemize}
\item{
  {\tt startFortMethod:sender}, the action of the Run button, which
	calls the Fortran subroutine {\tt startsub}.}
\item{
  {\tt continueFortMethod:sender}, the action of the Continue button,
	which calls subroutine {\tt continuesub}.}
\item{
  {\tt finishFortMethod:sender}, the action of the Finish Up button,
	which calls subroutine {\tt finishsub}.}
\item{
  {\tt clearForNewRun:sender}, the action of the Clear button, which simply
	calls the {\tt clearNXYView} method provided by {\tt nxyView} and
	clears the output Form displays.}
\end{itemize}
%
\noindent
Three methods are need for plotting graphs (Fig. \ref{plotmeth}):
\begin{itemize}
\item{
  {\tt updateOutputData}, which calls {\tt writeplotdata}.}
\item{
  {\tt sendPlotDataToWindow}, which also checks that the data is plottable.}
\item{
  {\tt provideDataStream}, which tells {\tt nxyView} where to find the data
  to plot.  ({\tt RunFortObjectInstance} is {\tt nxyView}'s ``delegate'',
  which means that for some messages received, {\tt nxyView} asks its
  delegate to respond.)}
\end{itemize}
%
\noindent
Finally, there is one additional method (not mentioned in the header file,
{i.e.}, not ``public''), {\tt displayIntmdteFs}, for handling the display
of intermediate results: the discontinuities, the calculated

changes to the scale parameters, and their present values.

The Fortran file that is associated with the {\tt rhoSky} application is

considerably larger than that needed for ``Hello, Fortran!''.
The communication with the Objective-C code continues to be, as above,
through a set of subroutine calls.
These subroutines were ``written'' by converting the
original Fortran main program (which itself largely consisted of subroutine
calls) into several top-level subroutines, already mentioned above with
regard to the methods connected with buttons:
{\tt startsub}, {\tt continuesub}, and {\tt finishsub}.
There are two additional top-level subroutines,
{\tt writeplotdata} (called by {\tt updateOutputData}) and
{\tt observables} (called by {\tt finishsub}).
The {\tt observables} subroutine originally was a

{\it separate} Fortran post-processor
program for calculating, from the final $F(r)$ and $G(r)$, the quantities
that eventually appear in the {\tt m\_rho}, {\tt M\_sky}, and {\tt GoldRat}
FormCells in the Main Window.
Finally, there are several lower-level routines needed for the Runge-Kutta
integration, such as {\tt DERIVS}, which calculates the right-hand-sides of the
four first-order equations.

What can {\it not} be shown in the figures of this article is the real-time
ease of use of the {\tt rhoSky} application.
Calculation proceeds quickly, with results and graphs appearing within a second
or so after clicking a button.
The interface in fact {\it encourages} a more exploratory approach to the
equations than one might undertake when searching for solutions
in a batch mode.
Most importantly, it is much more fun to ``drive'' this application

with the GUI interface than the old Fortran program with the command-line,
batch-mode interface.

\bigskip
\section{Portability of concept}

The interface described above is specific to one

particular physics problem, but it is
clear that the principles involved can be applied to many different problems
involving the solution of coupled differential equations.

In fact, we already have adapted (re-used) the code, with {\it very} little
extra work, to
solve some coupled equations for a quantum field theoretic

problem involving pions and $\sigma$-mesons.
This conversion only took about eight hours and most of that time was spent in
writing the prologue (i.e., comment lines)

which describes the problem being solved.
Essentially, only the Fortran subroutines {\tt LOAD1}, {\tt LOAD2}, and {\tt
DERIVS}, which depend upon the equations being solved,

had to be changed.
There were also some problem-specific changes
that had to be made, using the Interface Builder,

to the input and output Forms of the Main Window
(different coupling constants, different numbers of
functions to be solved for, etc.), but these were easy to make

using the drag-and-drop and editing capabilities of the Interface Builder.

In my experience as a programmer, this rapid turnaround in a programming

cycle, from original conception to the production of useful results, was

quite remarkable.
The exercise was very convincing, to me at least, that the ease of re-use

of object-oriented code modules is a real gain for the scientific programmer,

not just salesman's hype.

\bigskip
\nonum
\section{Acknowledgements}

This work evolved from research done in collaboration with Michael Mattis
and James L. Hughes.
I want to thank Bryan Travis, Jeremy Brackbill, James Gubernatus, and Klaus
Lackner for useful discussions regarding the interface described here and what
other projects could be approached with it.
An important component of the {\tt rhoSky} interface has been the
{\tt nxyPalette} object which displays the graphs in the Plot Window;
its author, Charles Fletcher of Techno-Sciences, Inc.,
was very helpful in getting it to work.
Finally, Kim Maltman, Juan P\'erez-Mercader and an unknown referee

made useful comments for improving the manuscript.

\newpage
\nonum
\section{Appendix: the Model and its Equations}

The {\tt rhoSky} application calculates the classical pion
and $\rho$-meson field functions for a $\rho$-stablized
skyrmion.\cite{rhoModel}
This is done using the so-called ``Hedgehog Ansatz'', which assumes the field
function solutions
are particular tensor covariants times spherically symmetric functions.
In this case, since there are two
meson fields, there are two functions, $F(r)$ and $G(r)$.

Mathematically, we want to solve two nonlinear differential equations for
$F$ and $G$ knowing the boundary conditions and
indicial behavior at $r=0$ and appropriate asymptotic
behavior (exponential damping).
The equations are
%
\begin{eqnarray}
r^2 F'' + 2rF' +(a-1)\sin{2F} \,\quad \quad \quad && \nonumber \\
  + 2a(G-1)\sin{F}- m_\pi^2 r^2 \sin{F} &\;=\;& 0 \ ,\\
r^2 G'' - (m_\rho^2 r^2 + 2)G + 3G^2 \quad \quad \quad && \nonumber \\
   - G^3 + m_\rho^2 r^2 (1-\cos{F}) &\;=\;& 0 \ .
\end{eqnarray}
%
Near $r=0$ the solutions must behave as
\begin{equation}
F = \pi - Ar + O(r^3) \ , ~~~ G = 2 - Br^2 + O(r^4) \ ,
\end{equation}
where the scale parameters $A$ and $B$ are constants to be determined by the
nonlinearity of the equations themselves.
Similarly, the desired asymptotic behavior at large $r$ is
\begin{eqnarray}
 F(r) &\rightarrow& C {\exp(-m_\pi r) \over r} \ ,\\
 G(r) &\rightarrow& {m_\rho^2 \over (m_\rho^2 -  4m_\pi^2)} {F^2(r) \over 2}
                    + D \exp(-m_\rho r) \ ,
\end{eqnarray}
where $C$ and $D$ are the other two scale parameters (constants) to be
determined by solving the nonlinear equations.

Note that, since $\rho$ couples to two pions, $G(r)$ falls off like

$\exp(- 2 m_\pi r)$, rather than $\exp(-m_\rho r)$.

\newpage

\newpage
\widetext

\figure{The interface window for the Hello, Fortran! application.  As it
appears, the user has just entered a number in the ``In'' FormCell and
is about to click on the Run button to initiate the Fortran calculation.}

\figure{The header (or interface)
file, {\tt runFortObject.h}, as generated by the NeXTSTEP
Interface Builder program after clicking on the Parse operation.  No further
code needs to be added to this file, which defines the outlets and methods
of the {\tt runFortObject} class, a custom-built subclass of the generic

Object class.}

\figure{The implementation file {\tt runFortObject.m}, the shell of which

is also generated by the Interface Builder application.  In this case the
programmer must flesh out the code so that the application

actually does something.  Here, the method {\tt runFortMethod:sender}
consisted only of the ``{\tt return self;}'' line before it was filled out.
Note that the method calls a C function, {\tt hellosub\_()}, which corresponds
to the Fortran subroutine {\tt hellosub} defined in file {\tt hellosub.f}.}

\figure{The Fortran file {\tt hellosub.f}.  This defines a subroutine
which reads input from a file generated by {\tt runFortMethod},
massages it, and then writes the output to a file which will be
accessed by {\tt runFortMethod} when {\tt hellosub} returns.}

\figure{The files {\tt make.preamble} and {\tt make.postamble} for the
Hello Fortran! application.  The preamble specifies the f2c library (needed for
the Fortran I/O routines) and declares {\tt hellosub.o} as an object file
to be linked into the application.  The postamble tells Make how to compile
{\tt hellosub.o} from its Fortran source.  Here {\tt f77} is a script that

invokes the f2c translater/compiler.}

\figure{The {\tt rhoSky} Main Window.  Input parameters are set in the
FormCells to the left and the starting values of the scale parameters,
A...D, are set in the upper middle set of cells.  The F's on the upper

right give the discontinuities of the functions $F(r)$ and $G(r)$ and their
derivatives after each pass (Runge-Kutta integration).  The changes in the
scale parameters calculated by a Newton-Raphson method to drive those
discontinuties to zero are shown in the lower right cells.  At each pass
the present values of the scale parameters appear in the middle-center

group of cells.  Finally, after achieving a good solution, clicking on

Finish Up invokes the computations whose results are shown in the lower
middle group of cells.}

\figure{The {\tt rhoSky} Plot Window, which appears on the screen along with
the Main Window showing input and output values as the iteration proceeds.
Shown are a typical first pass plot, with big discontinuities at the match

point (here at 0.5 fm), and after a solution has been attained.
\label{plotwin}}

\figure{The {\tt updateOutputData}, {\tt send\-Plot\-Data\-To\-Window},
and {\tt provide\-Data\-Stream} methods needed to drive the plotting object
in the {\tt rhoSky} Plot Window.
\label{plotmeth}}


\begin{references}

\bibitem [1] {NumRep}
  W. H. Press, B.\ P.\ Flannery, S.\ A.\ Teukolsky, and W.\ T.\ Veterling, {\it
Numerical Recipes: The Art of Scientific Computing} (Cambridge U.\ P., New
York, 1986), esp. Chap. 15; see also
W. H. Press and S. A. Teukolsky, Computers in Physics {\bf 6}, 188-191 (1992).

\bibitem [2] {proc}
This ``two-boundary-value problem'' is solved by shooting out from the origin,
using an adaptive
Runge-Kutta routine, to some intermediate distance, $r_F$ ({e.g.}, 0.5 fm).
Then one shoots back to $r_F$ from a large value of $r$ ({e.g.}, 2 fm)

where the asymptotic behaviors of $F(r)$ and $G(r)$ are known.
The discontinuities at $r_F$ in $F$ and $G$ and their
derivatives then determine, using a generalized Newton-Raphson technique,
corrections to the initial scale parameters that should tend to drive the
discontinuities toward zero in the next iteration.
The code then goes on to make successive passes through the above shooting

procedure, iterating until it achieves a good solution.

\bibitem [3] {inpfile}
One could have chosen to have the input read from an editable input file, but
for present purposes it was preferable to query the user.

\bibitem [4] {f2c}
  From AT\&T, available via anonymous FTP in a
version specifically compiled for the NeXT from
{\tt sonata.\-cc.\-purdue.\-edu}
in {\tt /pub\-/next\-/2.0-release\-/binaries}.

\bibitem [5] {NeXTSTEP}
  NeXT, Inc., {\it NeXTSTEP Reference Manual}, Addison Wesley, ISBN
0-201-58136-1, 1992.
  Useful supplementary documents available by anonymous FTP from the
{\tt sonata} archive, Ref. 4, are:
  NeXT, Inc., {\tt NeXTSTEP\_Concepts}, in directory {\tt /pub\-/next\-/docs};
  M.\ Mahoney, {\tt IB\_tutorial}, in
{\tt /pub\-/next\-/Newsletters\-/SCaNeWS};
  and J.\ Glover, {\it Short-Course on Object-Oriented Programming},
{\tt UHOOP\-class}, in {\tt /pub/next/docs}.

\bibitem [6] {nxyPal}
  C.\ Fletcher, {\tt nxyPalette1.2}, available from {\tt sonata}, Ref. 4, in
  directory {\tt /pub\-/next\-/submissions}.

\bibitem [7] {nxyPlot}
  D.\ Jesperson and T.\ Pulliam, {\tt nxyPlot1.8}, available from {\tt sonata},
  Ref. 4, in directory {\tt /pub/next/submissions}.

\bibitem [8] {rhoModel}
  Y.\ Igarashi et al., Nucl.\ Phys.\  {\bf B259}, 721-729, 1985.
  This model has a defect: the skyrmion solution is not really stable.  See
  Z.\ F.\ Ezawa and T.\ Yanagida, Phys.\ Rev.\ D {\bf 33}, 237, 1986 and
  J.\ Kunz and D.\ Masak, Phys. Lett. {\bf B179}, 146-152, 1986.
  (My colleagues and I only learned of this defect, of course, after the
  {\tt rhoSky} application was built and working.)


\end{references}
\end{document}